\documentclass[twocolumn,showpacs,preprintnumbers,amsmath,amssymb]{revtex4}
\usepackage{graphicx}
\usepackage{dcolumn}
\usepackage{bm}

\newcommand{\be}{\begin{eqnarray}}
\newcommand{\ee}{\end{eqnarray}}

\newcommand{\la}{\langle}
\newcommand{\ra}{\rangle}

\begin{document}

\title{Effects of disorder on lattice Ginzburg-Landau model of \\
$d$-wave superconductors and superfluids}

\author{Tomonori Shimizu$^\ast$} 
\author{Shunsuke Doi$^\dag$}
\author{Ikuo Ichinose$^\ast$}
\author{Tetsuo Matsui$^\dag$}
 \affiliation{${}^\ast$Department of Applied Physics, Graduate School of 
Engineering, \\
Nagoya Institute of Technology, 
Nagoya, 466-8555 Japan 
}
\affiliation{%
${}^\dag$Department of Physics, Kinki University, 
Higashi-Osaka, 577-8502 Japan
}%

\date{\today}

\begin{abstract}
We study the effects of quenched disorder on 
the two-dimensional   
$d$-wave superconductors (SC's) at zero temperature 
by Monte-Carlo simulations.
The model is defined on the three-dimesional (3D) lattice
and  the SC pair field is put on 
each spatial  link as motivated
in the resonating-valence-bond theory of 
the high-$T_{\rm c}$ SC's.
For the nonrandom case, the model exhibits 
a second-order phase transition to a SC state as density of 
charge carriers is increased.
It belongs to the universality class {\it different from}
that of the 3D XY model.
Quenched disorders (impurities) are introduced both in  
the hopping amplitude 
and the plaquette term of pair fields.
Then the second-order transition disappears 
at a critical concentration of quenched disorder,
$p_c\simeq 15\%$.
Implication of the results to cold atomic systems in  optical
lattices is also discussed.

\end{abstract}

\maketitle
Effects of disorder on phase structures and phase transitions 
have been studied for various systems.
In particular, the high-$T_{\rm c}$ materials are 
microscopically highly nonuniform and it is suggested that
there exists a spinglass like phase near the phase 
transition point of superconductivity (SC) at low temperatures 
($T$)\cite{SG}.
Furthermore, existence of a Bose glass was recently suggested  
in the Mott-insulating phase of cold atomic systems in random 
potentials\cite{Boseglass}.
In the present paper,  being motivated in part 
by these observations,
we shall study effects of quenched disorders on  SC's
by using a lattice Ginzburg-Landau (GL) model of unconventional
$d$-wave SC's.
The model in its pure case was
introduced as the GL theory of 
the resonating-valence-bond (RVB) field for the t-J model\cite{nm},
and the case of including interaction with the electromagnetic 
(EM) field
has been investigated recently\cite{smi}.
We expect that this model also describes superfluid phase of
fermionic atoms in cold atomic systems in optical lattices.
There, the effects of disorders can be investigated well under control.

We are interested in quantum SC phase transition 
of the two-dimensional (2D) model at $T=0$.
The model in path-integral representation is described 
in terms of the following RVB-type Cooper pair field $U_{xj}$:
\be
U_{xj}\sim \langle \psi_{x+j,\uparrow}\psi_{x\downarrow}
-\psi_{x+j,\downarrow}\psi_{x\uparrow} \rangle,
\ee
where $x(x_0,x_1,x_2)$ is the 
site of the 3D cubic lattice of size $V=L^3$
with periodic boundary condition and 
$j=1,2$ denotes
the spatial direction and also the unit vector in 
$j$-th direction.
$\psi_{x\sigma}$ is the electron annihilation operator at site
$x$ and spin $\sigma=\uparrow,\; \downarrow$.
In the $d_{x^2-y^2}$-wave SC, the Cooper pair amplitudes
in $x=x_1$ and $y=x_2$ 
have the opposite signatures as
$\langle U_{x1}U^\dagger_{x2} \rangle <0$.

Below we shall neglect the effects of the EM field 
and focus on the effects of quenched disorders.
The action of the clean system 
is then given as follows:
\begin{eqnarray}
A &=&g\sum_x\Big[ 
 c_{2}\: \bar{U}_{x2}U_{x+2,1}\bar{U}_{x+1,2}U_{x1}  \nonumber  
\ee
\be 
 &&+c_3\: (\bar{U}_{x+1,2}U_{x1}+U_{x+2,1}\bar{U}_{x+1,2} \nonumber \\
 &&\ \ \ \  +\bar{U}_{x2}U_{x+2,1}+U_{x1}\bar{U}_{x2})  \nonumber  \\
 &&+c_4\: (\bar{U}_{x+2,1}U_{x1}+\bar{U}_{x+1,2}U_{x2}
 )  \nonumber  \\
 &&+c_5\: (\bar{U}_{x+0,1}U_{x1}+\bar{U}_{x+0,2}U_{x2}
 )+\mbox{c.c.} \Big],
\label{action}
\end{eqnarray}
where we consider the London limit and set $U_{xj}$
a U(1) variable, $U_{xj}
=\exp(i\theta_{xj}),\ \theta_{xj} \in [-\pi, \pi)$.
The overall factor $g$ plays a role of $1/\hbar$ and
controls {\em quantum fluctuations}.
When we fix $c_i$, $g$ can be taken as an increasing function of 
the carrier concentration $\delta$.
(In the t-J model $\delta$ is the hole density.)
The $c_2$ terms controls fluctuation of flux
of $U_{xj}$'s around each spatial plaquette.
The $c_3$ and $c_4$ terms represent the spatial hopping of $U_{xj}$,
whereas the $c_5$ term describes the hopping in the 
imaginary-time ($0$-th) direction.
The partition function $Z$ is given by $Z=\int[dU]\exp(A),\
[dU]=\prod_{x,j}d\theta_{xj}/(2\pi)$.
We consider the parameter region $c_3 <0$ to expect 
$\langle U_{x1}U^\dagger_{x2} \rangle <0$, 
although $Z(c_3) = Z(-c_3)$ because of the change of variables
$U_{x1} \rightarrow -U_{x1}$.
The action $A$ is related to the 
action $A_{\rm Higgs}$ 
of the U(1) Higgs gauge theory 
that is obtained from Eq.(\ref{action})
by the replacement $U_{xj}\rightarrow \bar{\phi}_{x+j}U_{xj}
\phi_x$, where $\phi_x = \exp(i\varphi_x)$ is the U(1) Higgs field.
$A_{\rm Higgs}$ is invariant
under time-independent local gauge transformation
$\varphi_x\rightarrow \varphi_x+\lambda_{x},\
\theta_{xj}\rightarrow \theta_{xj}+\lambda_{x+j}-\lambda_{x}$
where $\lambda_{x(x_1,x_2)}$ is an $x_0$-independent function.
Actually, $A$ is viewed as the gauge-fixed version of $A_{\rm Higgs}$
in the unitary gauge $\phi_x = 1$. 

Queched disorder is introduced in the system (\ref{action})
by replacing the coefficinets $c_{2,3,4}$ 
as spatial-plaquette dependent ones.
First, we consider the 2D spatial plane with fixed $x_0$,
say $x_0=0$. Among $L^2$ plaquettes in the plane 
we choose $p\times L^2$  plaquettes randomly 
as ones at which impurities reside. We call it a sample.
We consider that the configuration of ``wrong" plaquettes
is $x_0$-independent because the location of impurities
are fixed along the imaginary time.
Then we reverse the values of $c_{2,3,4}$
for interaction terms contained in these plaquettes\cite{disorder}.
Thus the new plaquette-dependent coefficients $c^p_{2,3,4}$
are given by
\begin{equation}
c^p_{2,3,4}=
\left\{
  \begin{array}{cl}
  c_{2,3,4} & \mbox{with\ probability $1-p$}  \\
  -c_{2,3,4} & \mbox{with\ probability $p$},
  \end{array}
\right.
\end{equation}  
{\em for each spatial plaquette}.
Please note $c^p_5=c_5$.
Then the partition function $Z_p$ 
and the free energy
$F_p$ per site of one sample are given by
\be
Z_p=\int [dU]\exp(A_p)=\exp(-VF_p),
\label{FZ}
\ee
where $A_p$ is obtained from Eq.(\ref{action})
by replacing $c_i$ by $c^p_i$.
To obtain an ensemble average $\la O \ra$ of an observable $O(\{U_{xj}\})$
in the disordered system, we first prepare $N_s$ samples
and calculate the quantum-mechanical average $\la O \ra_s$ for 
the $s$-th sample ($s=1, \cdots, N_s)$.
Then we average it over samples,
\be
\la O \ra=\frac{1}{N_s}\sum_{s=1}^{N_s}
\la O \ra_s, \la O \ra_s =Z_{p_s}^{-1}\int[dU]O\exp(A_{p_s}).
\ee

For the MC simulations, we used the standard Metropolis 
algorithm\cite{ma}.
The typical statistics used was $10^5$ sweeps per block and the
average and MC errors were estimated over 10 blocks 
for each sample. Then we take
quenched averages over $N_s = 30\sim 50$ samples.
We estimated standard deviation  of physical quantities like ``specific heat" 
over samples (we call it sample error) as a function of $N_s$, which becomes
stable for $N_s \stackrel{>}{\sim} 30$.

\begin{figure}[t]
\includegraphics[width=6.5cm]{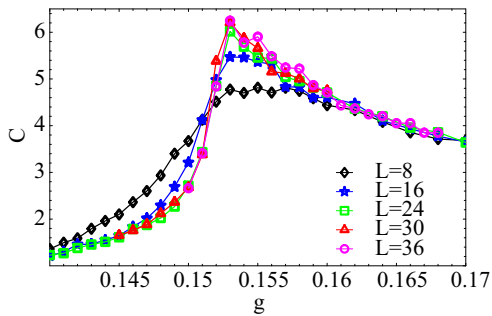}
\caption{\label{fig1}
Specific heat $C$ for $p=0$ as a function of $g$ with
$c_2=-c_3=c_4=c_5=1$. Error bars represent MC errors.
System-size dependence of its peak supports existence of a second-order phase transition.
}\end{figure}

\begin{figure}[t]
\includegraphics[width=6cm]{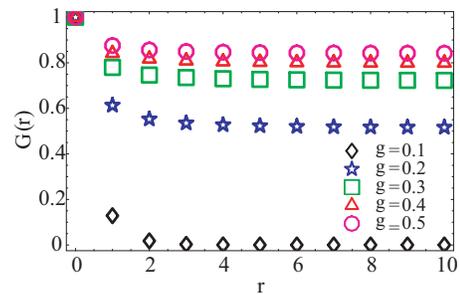}
\caption{\label{fig2}
Correlation function $G(r)$ for various values of $g$ with
$c_{2,4,5}=-c_3=1$.
Critical coupling $g_c$ is estimated $g_c=0.154$ for $L=20$.
Data show that for $g>g_c$ there exists SC LRO, 
whereas $g<g_c$ no LRO.
}\end{figure}


Let us first consider the nonrandom case, $p=0$.
We studied the phase structure by calculating the 
``internal energy"
per site $E=-\langle A \rangle/V$ 
and the ``specific 
heat" per site $C=\langle (A-\langle A\rangle)^2
\rangle/V$.
Typical behavior of $C$ is shown in Fig.\ref{fig1}, which indicates 
 a second-order phase transition
at $g = g_c \simeq 0.154$.
This transition has been predicted in Ref.\cite{nm} 
and also observed in the presence of
the EM gauge interactions\cite{smi}.
In order to verify that it is a transition to a SC phase,
we measured the correlation function of $U_{xj}$,
the order parameter field of SC,
\begin{equation}
G(r)={1 \over 4V}\sum_{x,i\neq j}
\langle \bar{U}_{xj}U_{x+ri,j} \rangle +\mbox{c.c.}
\label{Gr}
\end{equation}
We present $G(r)$ in Fig.\ref{fig2}.
It is obvious that there exists SC long-range order (LRO) 
for $g>g_c$, whereas  no LRO for $g<g_c$\cite{lro}.
Therefore the observed transition is nothing but 
the SC phase transition.

The critical exponents for $c_{2,4,5}=-c_3=1$ were estimated by 
the finite-size scaling analysis for $C$.
We obtained $\nu=1.5,\; \alpha=0.285$ and the critical coupling
$g_\infty=0.153$.
When $c_2=c_3=0$, the system (\ref{action}) reduces to a
set of decoupled 2D XY spin models 
[$U_{xj}$ plays the role of an XY spin], 
each of which has the Kosterlitz-Thouless transition.
The $c_2$ and $c_3$ terms couple these 2D XY spins.
From the above values of critical exponents, we judge that the present
phase transition does not
belong to the universality class of the 3D XY model
having $\nu= 0.6721(13)$\cite{exponent}.

\begin{figure}[b]
\includegraphics[width=6.5cm]{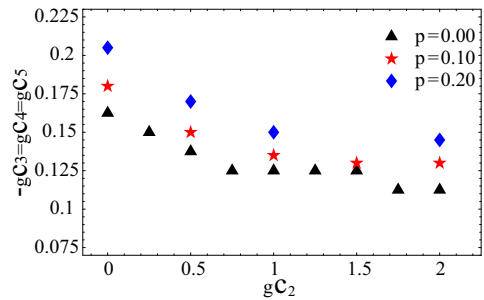}
\caption{\label{fig3}
Location of the peak of $C$ with $L=12$ and $N_s=30$
in the $g c_2 - g\times(-c_3=c_4=c_5)$
plane for $p=0,0.10,0.20$. The region of disorder ``phase"
increases as $p$ increases.
}\end{figure}

Let us next turn to the random case, $p > 0$.
At first, we consider the $g c_2- g c'$ plane
where $c'\equiv -c_3 = c_4 = c_5$, and search for 
  the location of the peak of $C$. 
In Fig.\ref{fig3}, we present the peak location of $C$
for given disorder concentration
$p = 0.0, 0.10, 0.20.$
For $p=0.0$, the peak line expresses the second-order transition
as we saw in Figs.\ref{fig1} and \ref{fig2}.
We see that the region of the normal (non SC) state
increases as $p$ increases, although we need to
check whether the location of this peak expresses genuine
SC transition for $ p > 0$.  Below we examine it 
\noindent
by focusing on the specific case $c_2=-c_3=c_4=c_5=1$ with 
the varying parameter $g$ and calculate
$E,\ C,\  G(r)$, etc.
Density of quenched disorder that we studied is 
$p=0.05, 0.10, 0.15$ and $0.20$. 

\begin{figure}[t]
\includegraphics[width=7cm]{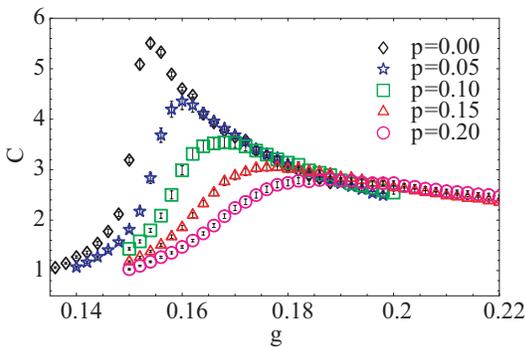}
\caption{\label{fig4}
Specific heat $C$ for $L=16$ with $N_s = 30$
vs $g$ for $p=0.0, \cdots, 0.20$.
Error bars here and below represent sample errors.
The peak around the SC phase transition 
is getting weaker as $p$ is increased.
Location of the peak moves to larger $g$.
}\end{figure}

We first present the result of $C$ in Fig.\ref{fig4}.
The signal of the second-oder phase transition at
$p=0$ is getting weaker as $p$ is increased, and also the location
of the peak moves to larger $g$.

\begin{figure}[b]
\includegraphics[width=7.5cm]{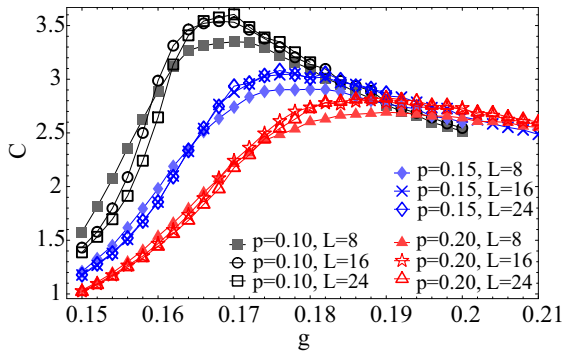}
\caption{\label{fig5}
The system size dependence of $C$ with $N_s =30$ 
for $p=0.10,\ 0.15,\ 0.20$. 
$C$ for $p=0.10$ shows SSD, 
which supports a second-order phase transition to SC state.
$C$ for $p=0.15$  has less SSD,
and $C$ for $p=0.20$ shows  no SSD for $L=16,24$.
}\end{figure}


\begin{figure}[t]
\begin{minipage}[t]{4.5cm}
\hspace{-1cm}
\includegraphics[width=4.1cm]{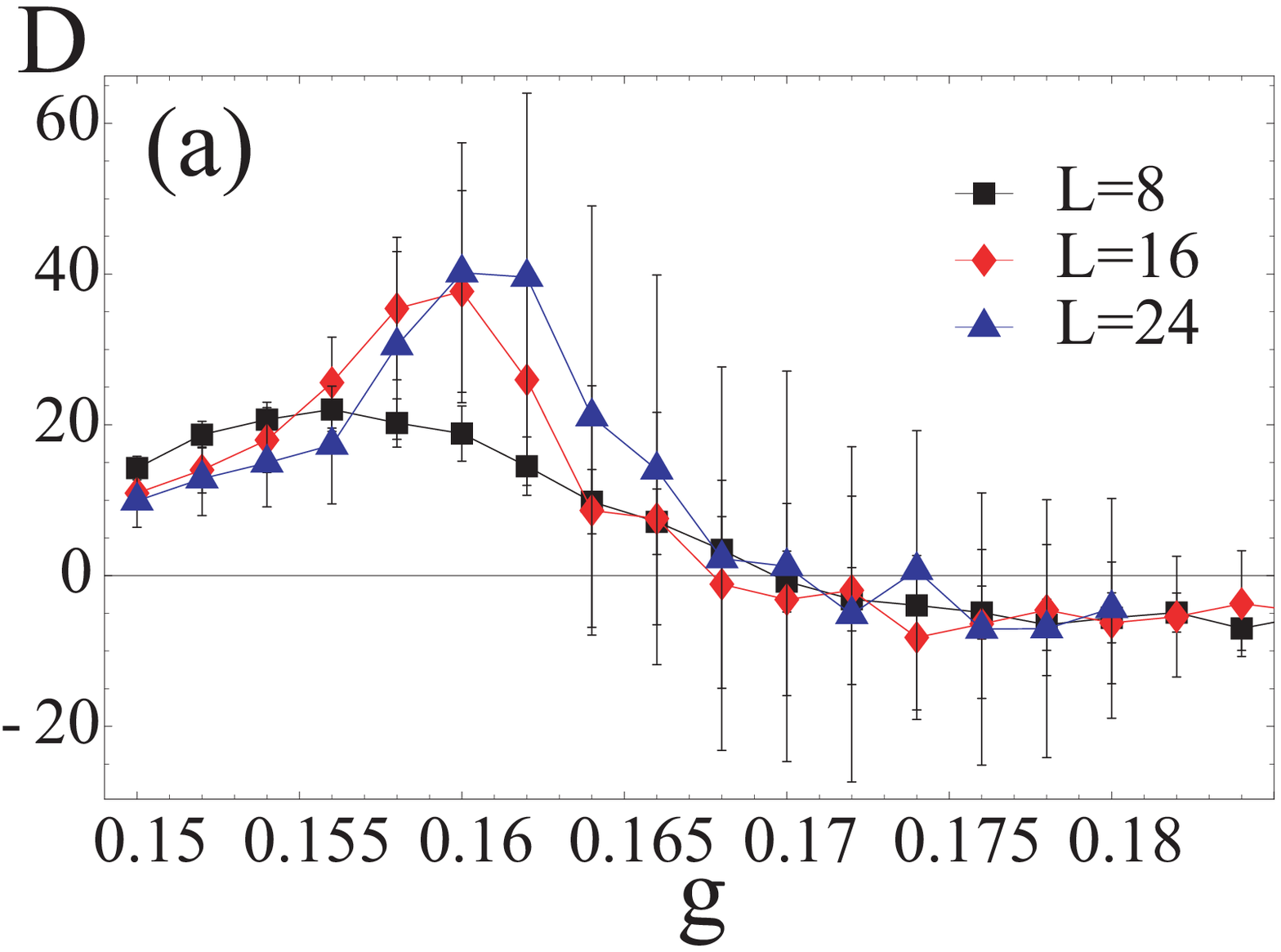}
\end{minipage}
\hspace{-0.8cm}
\begin{minipage}[t]{4.5cm}
\includegraphics[width=4.1cm]{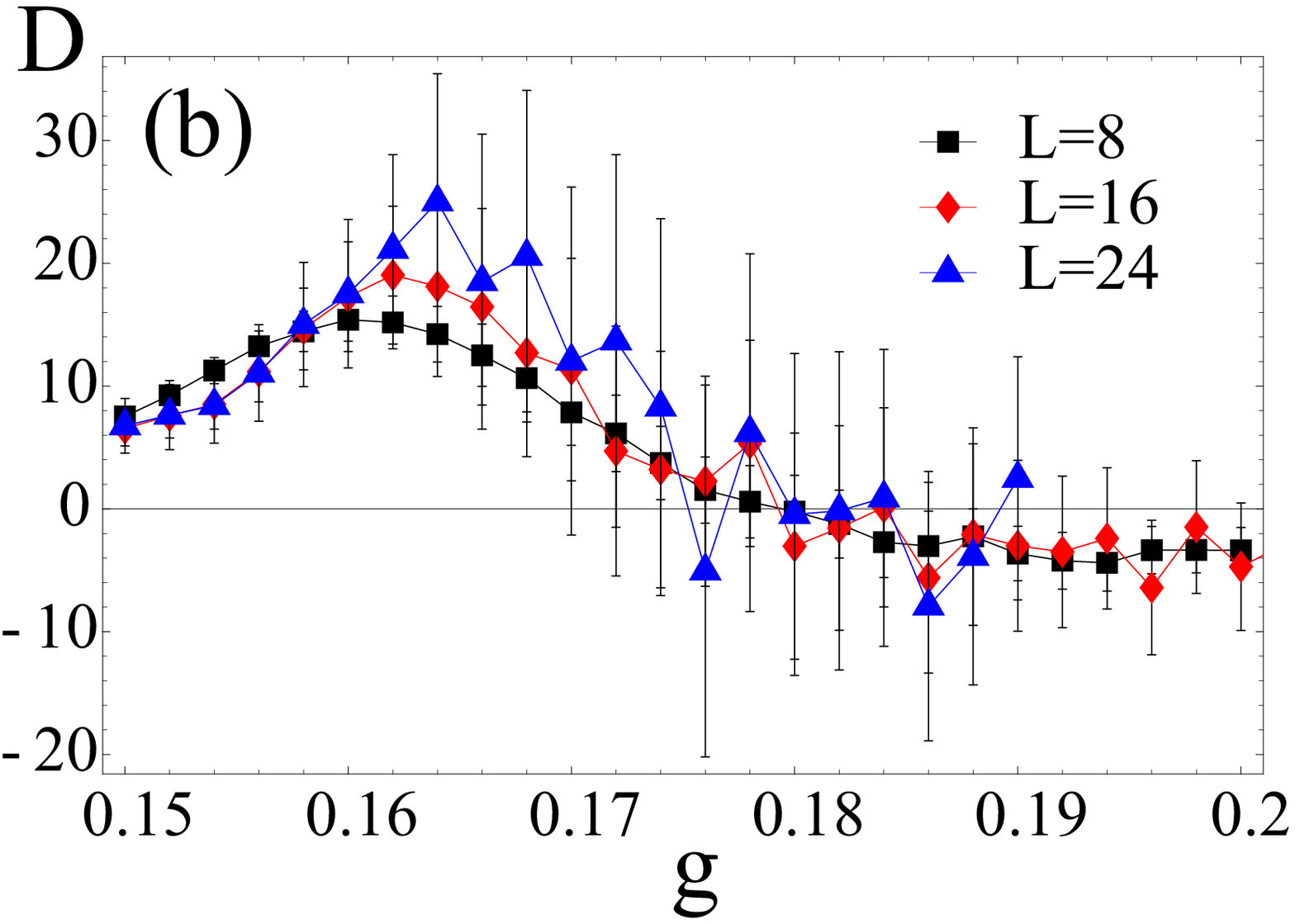}\\
\end{minipage}\\
\hspace{-1.cm}
\includegraphics[width=4.1cm]{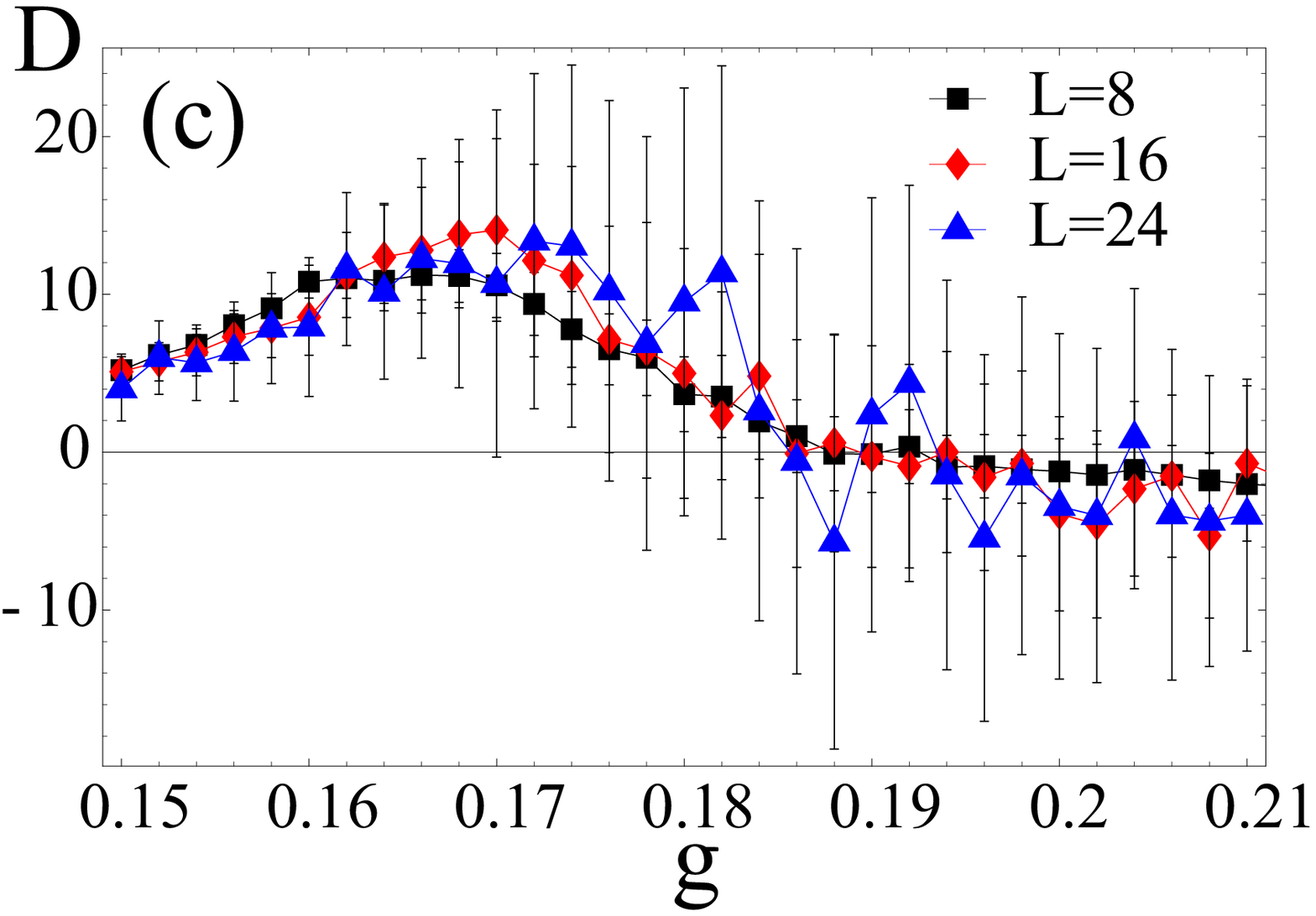}
\caption{\label{fig6}
$D(g) \equiv g dC(g)/dg$ with $N_s = 30$
vs $g$ for various $L$.
(a)$p=0.10$;\ (b)$p=0.15$;\
(c)$p=0.20$. The cases  (a) and (b)
show SSD, while the case (c) shows almost {\em no} SSD.
}\end{figure}


Next, in Fig.\ref{fig5},  we present 
the system-size dependence (SSD) of 
$C$ for $p=0.10, \ 0.15$ and $0.20$.
We also calculated the derivative of $C$,
$D(g) \equiv   g dC(g)/dg$, to identify the order of the phase transition.
Results are shown in Fig.\ref{fig6}.
From these results, we judge that there exists 
a second-order transition 
for $p\le 0.10$ whereas it changes to a crossover for $p\ge 0.20$.
The case of $p=0.15$ seems to have no SSD in $C$ but have certain 
SSD for $D(g)$.
Therefore, the transition in the $p=0.15$ case might be of third order.


\begin{figure}[b]
\begin{minipage}[t]{4.5cm}
\hspace{-1cm}
\includegraphics[width=4.2cm]{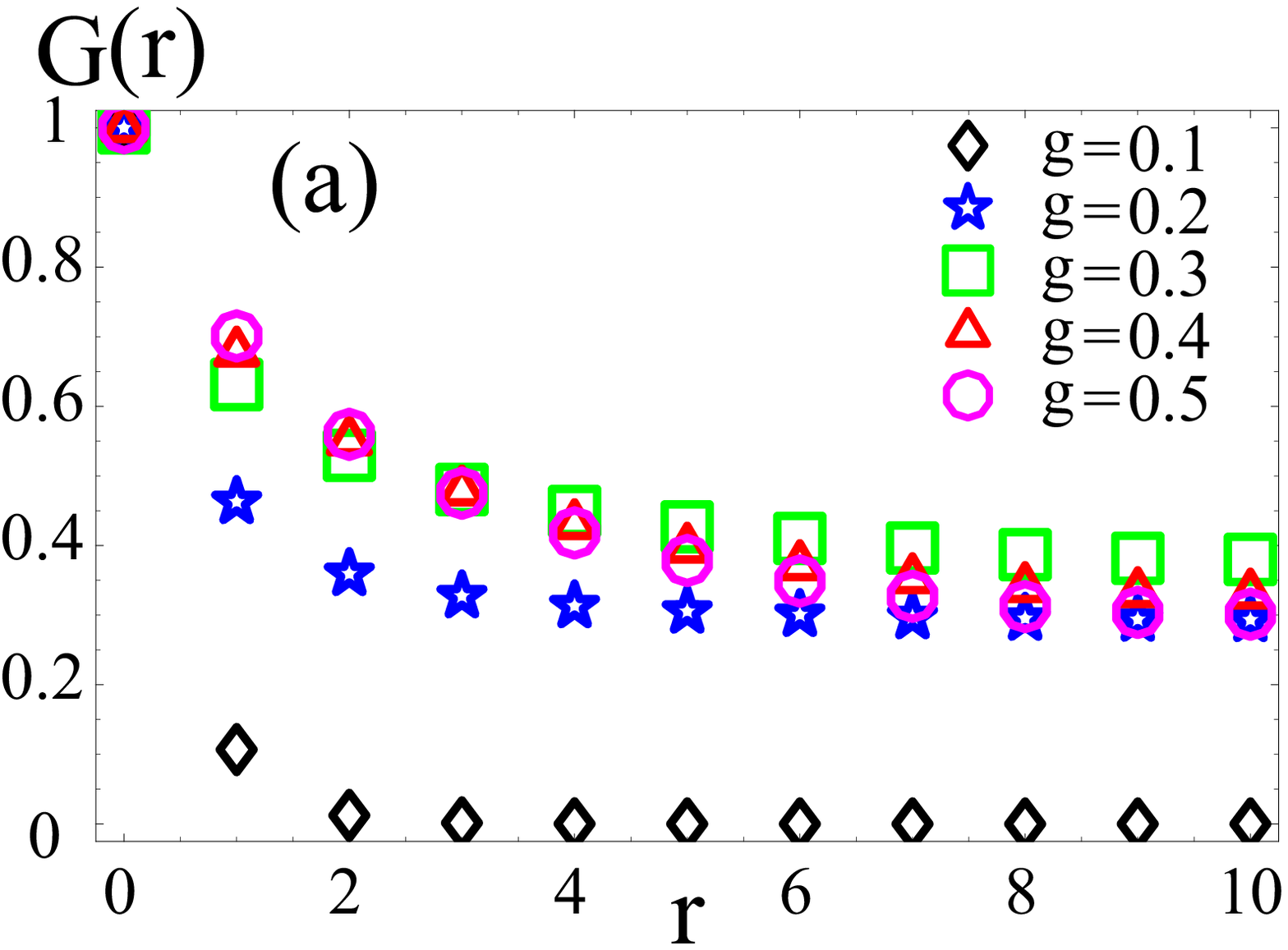}
\end{minipage}
\hspace{-1cm}
\begin{minipage}[t]{4.5cm}
\includegraphics[width=4.5cm]{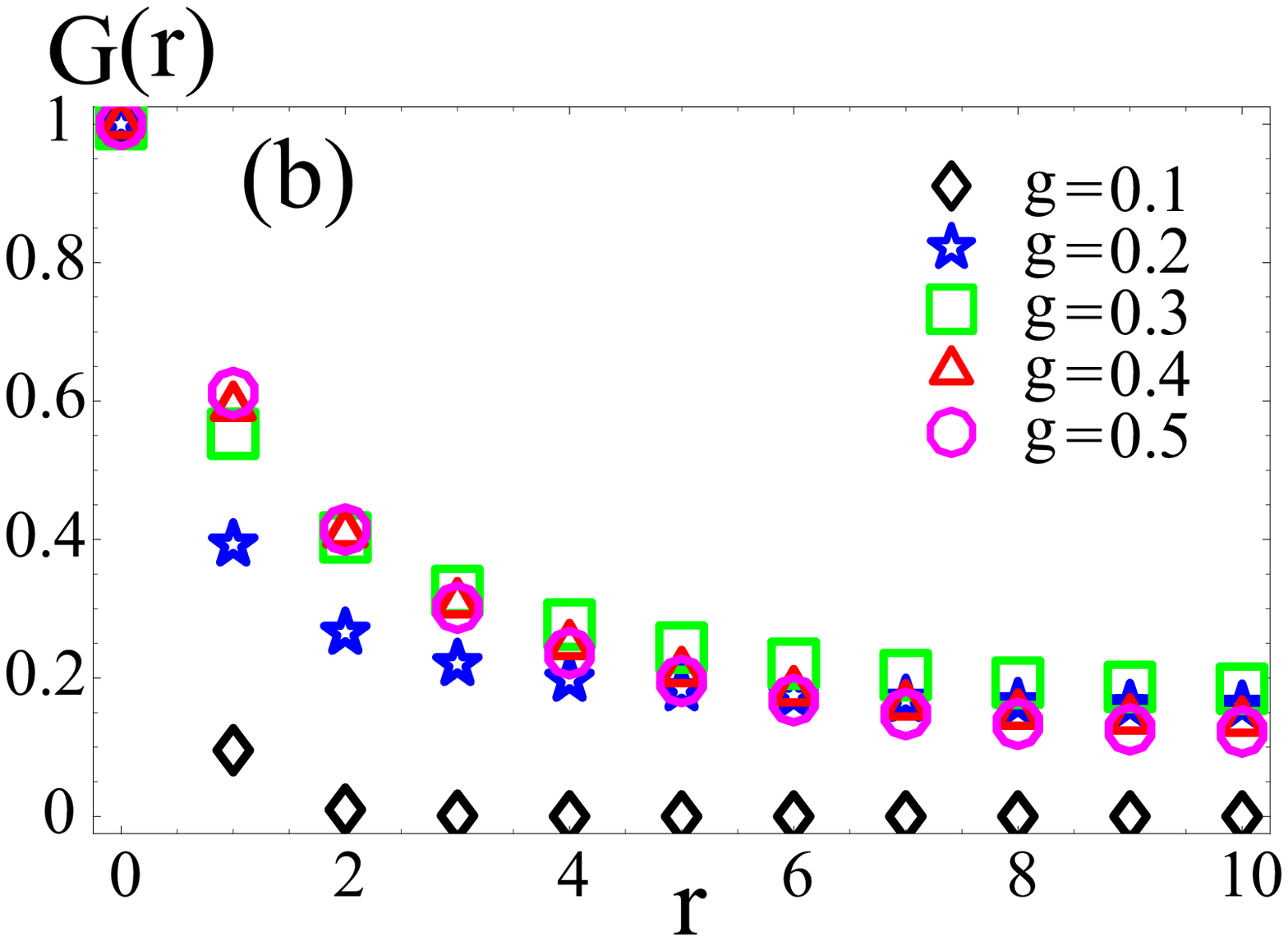}\\
\end{minipage}
\begin{minipage}[t]{4.5cm}
\hspace{-1.5cm}
\includegraphics[width=4.6cm]{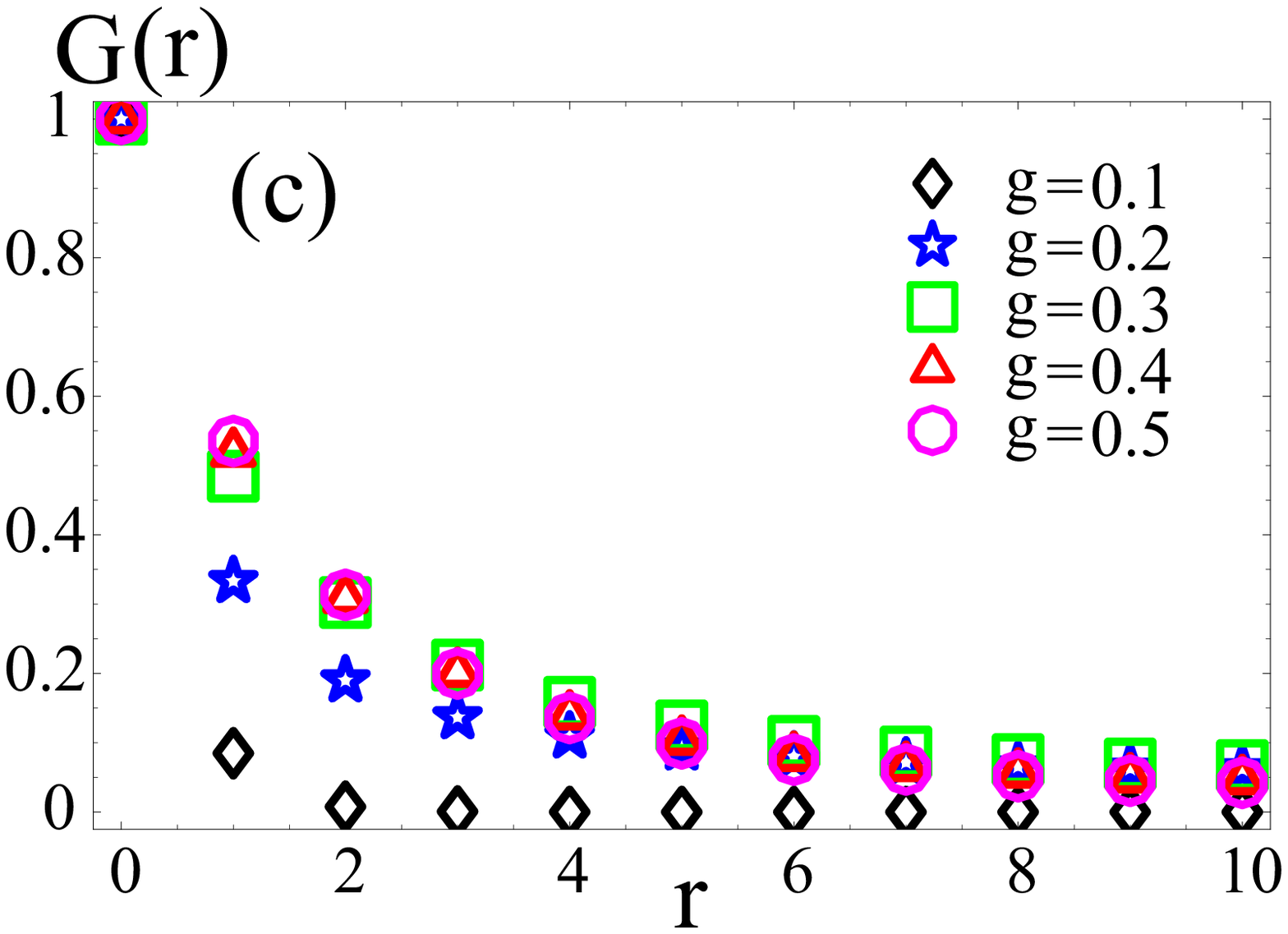}\\
\end{minipage}
\hspace{-1cm}
\begin{minipage}[t]{4.5cm}
\includegraphics[width=4.5cm]{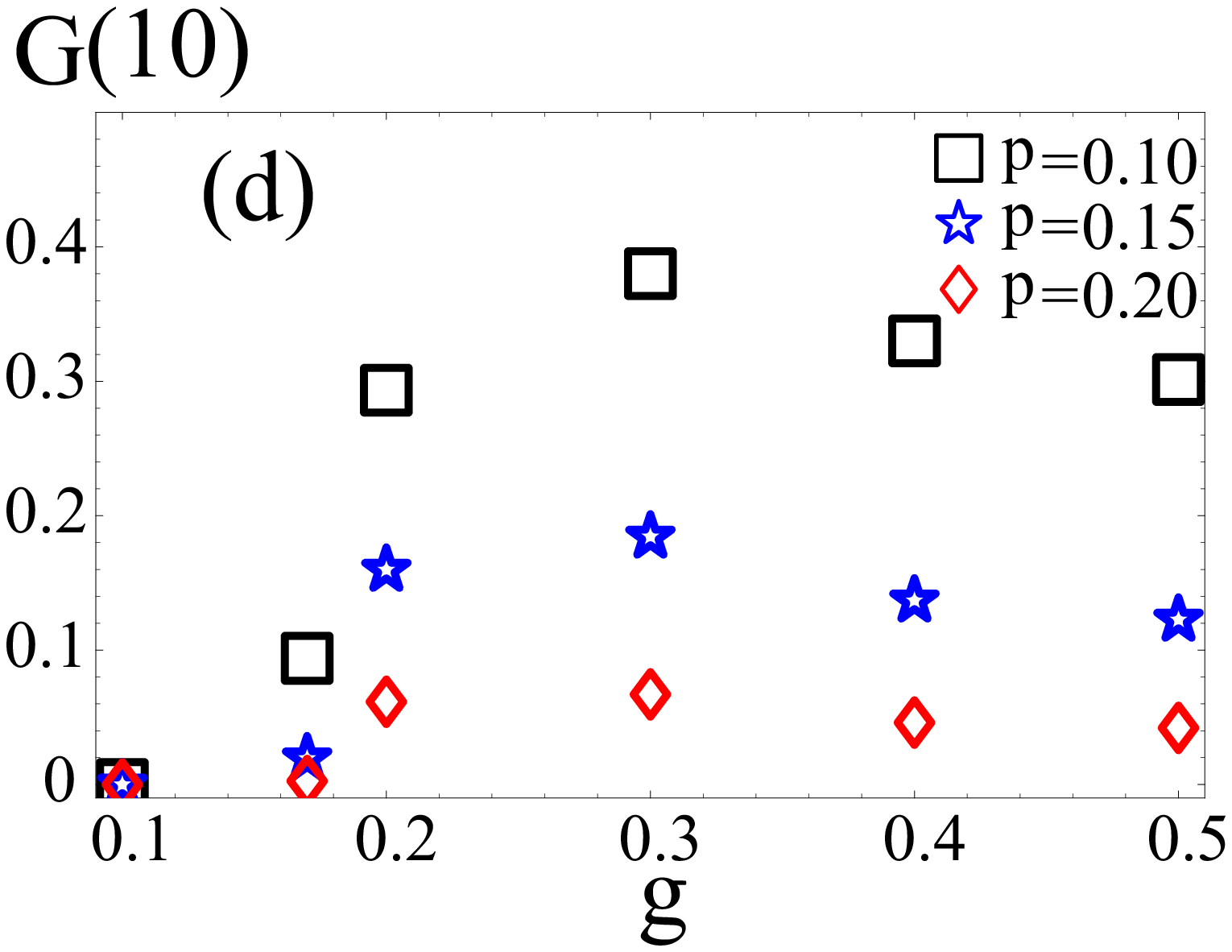}\\
\end{minipage}
\caption{\label{fig7}
SC correlation function $G(r)$ for $L=20$ with
$N_s=50$. (a)$p=0.10$;\ (b)$p=0.15$;\
(c)$p=0.20$. (d) $G(r=10)$ for (a,b,c) vs g.
SC ``transition point" $g_c$ is estimated from the specific heat $C$
as $g_c\simeq (a)0.168,\ (b)0.17,\ (c)0.18$. 
Fig.(d) supports that
LRO developes
 for $g>g_c$  in (a).
}\end{figure}

To study if the SC state persists even for $p \ge 0.20$,
we measured the SC correlation $G(r)$
averaged over samples with randomly generated $c^p_i$.
In Fig.\ref{fig7} we present the results for 
$p=0.10, \ 0.15$ and $0.20$. 
It supports that for $p=0.10$ there exists the SC LRO for 
$g>g_c$ whereas for $p=0.20$ no LRO for any values of $g$.
This observation leads to the conclusion that the SC phase
disappears for $p \ge 0.20$, i.e., there is a multicritical
point $p_{\rm mc}$ near $p=0.15$ in the $g-p$ phase
diagram.

\begin{figure}[t]
\includegraphics[width=7cm]{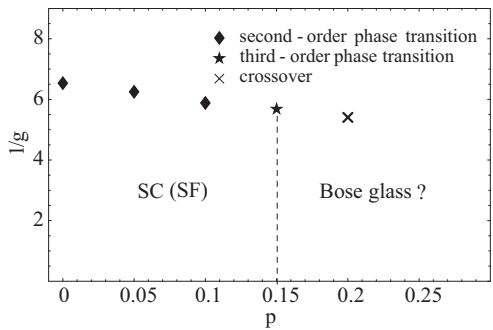}
\caption{\label{fig8}
Phase diagram in the $p-g^{-1}$ phase diagram
for the lattice model of dirty $d$-wave SC
with $c_2=-c_3=c_4=c_5=1$.
Possible Bose glass phase is suggested.
There should be a multicritical point near 
the line $p=0.15$. 
}\end{figure}


It is interesing to ask what kind of phase is realized
in the region $p>p_{\rm mc}$ and $g>g_{\rm cr}$, where
$g_{\rm cr}$ is the crossover coupling determined  by the
peak location of the specific heat.
Probably, in the case of the doped AF magnets 
with strong inhomogeneity, this phase may simply 
correspond to dirty ``normal" metallic state.
We think that the present model given by Eq.(\ref{action}) also
describes the superfluid $d$-wave RVB state of fermionic atoms
in 2D optical lattice, which was recently proposed by sereval 
authors\cite{dwave}.
Random disorders can be introduced into the systems
by a laser speckle or by an incommensurate bichromatic potentail.
In that experimental setup, there is an interesting possibility that a Bose 
glass phase, which is an analogue of the spin glass phase, is realized there.
For ultracold strongly interacting $^{87}$Rb bosons, 
a Bose glass phase is suggested by experiments\cite{Boseglass}.
In the experiments of cold atom systems in  optical lattices,
the Bose glass phase has no long-range coherence but excitations
are gapless.
In Fig.\ref{fig8}, an expected phase diagram of the present lattice
model of dirty $d$-wave SC and superfluidity (SF)  is shown.
We think that existence of the Bose glass phase is examined
theoretically  by the standard analytical and numerical methods 
applied for spin glass\cite{sg,z2}.



\begin{references}
\bibitem{lro}
Because we deal with the gauge fixed version of $A_{\rm Higgs}$
expectation values of  gauge-noninvariant quantities like
$G(r)$ may be  nonvanishing without conflicting
Elitzur's theorem.
\end{references}
\end{document}